\documentclass[a4paper,twoside]{article}

\usepackage{epsfig}
\usepackage{subfigure}
\usepackage{calc}
\usepackage{amssymb}
\usepackage{amstext} 
\usepackage{amsthm}
\usepackage{url}
\usepackage[table,xcdraw]{xcolor}
\usepackage[fleqn]{amsmath} 
\usepackage{pslatex}
\usepackage{enumitem}
\usepackage{textcomp}
\usepackage{natbib}
\usepackage{array,multirow,graphicx}

\usepackage{SCITEPRESS}     % Please add other packages that you may need BEFORE the SCITEPRESS.sty package.

\begin{document}

\title{
Easy Mobile Meter Reading for Non-Smart Meters:\\
Comparison of AWS Rekognition and Google Cloud Vision Approaches 
}

 
\author{ 
\authorname{ 
Maria Spichkova\sup{1}, Johan van Zyl\sup{1}, Siddharth Sachdev\sup{1}, Ashish Bhardwaj\sup{1},  Nirav Desai\sup{1}  
}
\affiliation{ 
\sup{1}School of Science, RMIT University\\
Melbourne, Australia\\
maria.spichkova@rmit.edu.au, $\{$s3235459,s3632346,s3615564,s3633257$\}$@student.rmit.edu.au
} 
}

\abstract{%
Electricity and gas meter reading is a time consuming task, which is done manually in most cases. There are some approaches proposing use of smart meters that report their readings automatically. However, this solution is expensive and requires (1) replacement of the existing meters, even when they are functional and new, and (2) large changes of the whole system dealing with the meter readings.  
This paper presents results of a project on automation of the meter reading process for the standard (non-smart) meters using computer vision techniques, focusing on the comparison of two computer vision  techniques, 
Google Cloud Vision  and AWS Rekognition. \footnote{Preprint. Accepted to the 14th International Conference on Evaluation of Novel Approaches to Software
		Engineering (ENASE 2019). Final version published by SCITEPRESS, http://www.scitepress.org}
}

\onecolumn 
\maketitle 
\normalsize \vfill

%===============================================
\section{\uppercase{Introduction}}
\label{sec:introduction}

There are many approaches proposing smart devices for
several types of utilities, see e.g., \cite{depuru2011smart, benzi2011electricity, zheng2013smart}. 
Smart meters can record energy consumption and automatically send the corresponding data to the electricity supplier for monitoring and billing purposes. This solution is definitely useful and has many benefits, from increasing the sustainability to offering potential benefits to householders. 
However, implementing it in a large scale in real life is expensive, e.g., the costs of the transition program for Australia were estimated to be a total cost of \$ 1.6 billions. 
Many customers prefer not to upgrade their non-smart meters to a smart version, when the costs of this upgrade are out of their pocket. 
For example, in Australia, different energy providers may have different approaches to how they charge their customers for this change -- either as a lump sum that is added to the first bill after the upgrade or a higher monthly fee -- but in all cases the costs are beared by the customers.  
Also, the use of smart meters raised privacy concerns from the consumers' side: as they typically record energy consumption on the hourly basis or even more frequently, and report it to the system at least daily, this information might be used to identify whether the  residences are at home or not, etc.

Therefore, many countries delay the transition to the smart meter systems or purpose a partial transition. Thus, another solution is required for this case, as to do the 
electricity and gas meter reading completely manually 
is extremely time consuming. 
%There are some approaches proposing use of smart meters that report their readings automatically. However, this solution is expensive and requires (1) replacement of the existing meters, even when they are functional and new, and (2) large changes of the whole system dealing with the meter readings. 
We proposed a solution for non-smart meters, which is based on computer vision approaches. This solution provides 
an easy way for customers to upload meter readings to their system. 

The project was conducted in collaboration with Energy Australia, which is an electricity and gas retailing private company in Australia. It supplies electricity and natural gas to more than 2.6 million residential and business customers throughout Australia. Our goal was to provide a convenient alternative method for their current meter reading updating system. The current solution from Energy Australia involves consumers using updating their utility reading through using an online portal.  
This method is inconvenient for consumers as consumers need to provide intricate entry details. Consumers are also required to calculate their utility reading from their meter. 
The proposed solution is to use a mobile application for capturing readings, a cloud-system to manage readings and a blockchain technology, see \cite{zheng2018blockchain, michael2018blockchain, swan2015blockchain}, to store reading securely.

\emph{Contributions:}  
This paper presents 
(1) the architecture and implementation details of the proposed solution, as well as 
(2) the comparison of two computer vision technologies, 
Google Cloud Vision\footnote{\url{https://cloud.google.com/vision}} and Amazon Web Services (AWS)  Rekognition\footnote{\url{https://aws.amazon.com/rekognition}}, applied for recognition in utility meter readings. As the majority of the currently used meters have digital displays (the old versions were of dial type)
we focused on this type of displays as well as on digit recognition analysis.

The system was elaborated within a research project under the initiative 
\emph{Research embedded in teaching}, proposed at the RMIT University, see \cite{spichkova2017autonomous,simic2016enhancing}.
To encourage curiosity of Bachelor and Master students to the research in Software Engineering, we suggested to include research and analysis components in the projects as a bonus task. 
Short research projects have been sponsored by industrial partners and focused on the topics related to the project to conduct within semester. These have to be conducted after the semester end, focusing on research prospective and deeper analysis of the semester task, see for example   \cite{ICECCS_SMI,gaikwad2019voice,chugh2019automated,sun2018software,spichkova2018automated,spichkova2016formal,christianto2018enhancing,clunne2017modelling}.

\begin{figure*}[ht!] 
\begin{center}
%\centering
\includegraphics[scale=0.9]{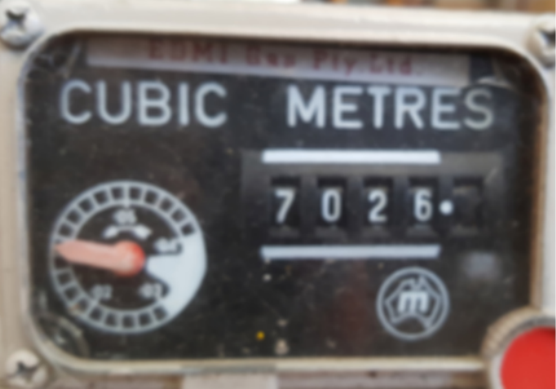}~\includegraphics[scale=0.9]{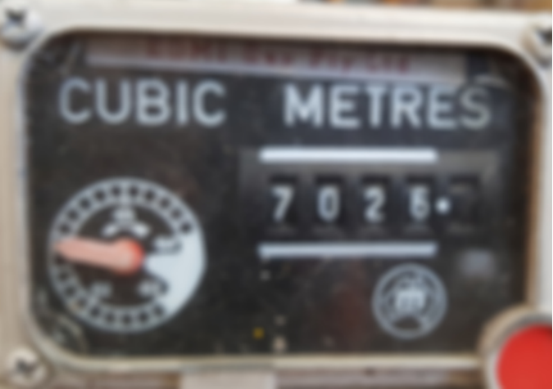}~\includegraphics[scale=0.9]{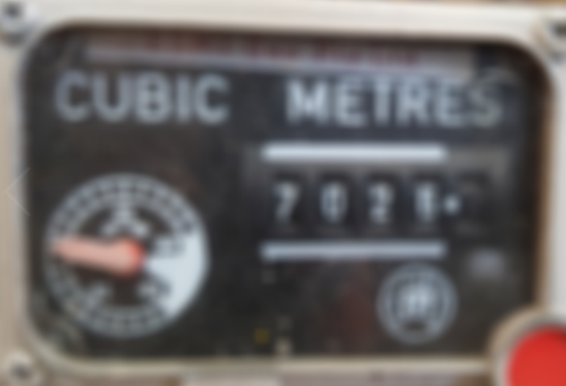}\\
(a) \hspace{5cm} (b) \hspace{5cm} (c)
\end{center}
\caption{Blurring effect: (a) 30BLUR, (b) 60BLUR, (c) 90BLUR}
\label{fig:blur}
\end{figure*}

\begin{figure*}[ht!] 
\begin{center}
%\centering
\includegraphics[scale=0.9]{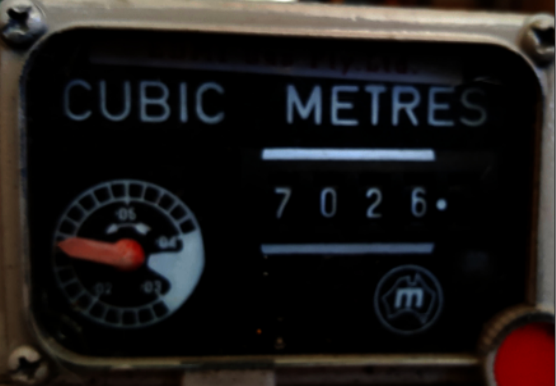}~\includegraphics[scale=0.9]{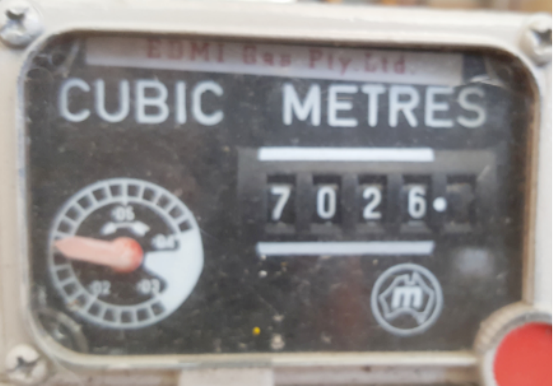}~\includegraphics[scale=0.9]{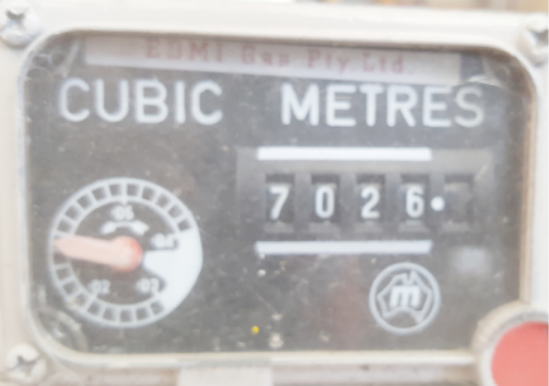}\\
(a) \hspace{5cm} (b) \hspace{5cm} (c)
\end{center}
\caption{Gamma correction effect: (a) 0.25GAMMA, (b) 1.5GAMMA, (c) 3.0GAMMA}
\label{fig:gamma}
\end{figure*}

\emph{Outline:} The rest of the paper is organised as follows.   
Section~\ref{sec:related} introduces related work.
The methodology we applied to compare AWS Rekognition and Google Cloud Vision technologies, as well as  the results of the conducted study are discussed in Section~\ref{sec:study}. The proposed and implemented system is presented in Section~\ref{sec:system}. 
Finally, Section~\ref{sec:conclusions} summarises the paper and proposed future work directions.  
% 

%===============================================
\section{\uppercase{Related Work}}
\label{sec:related}

Many research works on elaboration of automated/remote meter reading were conducted even approx. 20 years ago. There are also a number of corresponding patents. For example,  
\cite{kelley2000automated} patented an automated meter reading (AMR) system with distributed architecture that collects, loads, and manages system-wide data collected from energy meters and routes the data automatically to upstream business systems. 

\cite{nap2001automatic} patented an automatic meter reading data communication system that has an integrated digital encoder and two-way wireless transceiver that is attachable to a wide variety of utility meters for meter data collection and information management. 
Many other systems with similar ideas were patented  \cite{jenney1999automatic, knight1998remote, ehrke2003electronic}, but the research area is still very active, see e.g., \cite{grady2016method,winter2017methods}. 

However, the majority of works in this area last years focus on the following aspects:
\begin{itemize}
    \item 
    \emph{Application of the data mining and data analytics techniques on the meter reading data.} \\
    Thus, \cite{rathod2016regional} presented an electricity consumption analysis for consumers using data mining techniques applied to meter reading data. 
    \cite{xiao2013exploring} proposed an approaches to recognise energy theft based on the analysis of meter data. 
    \item
    \emph{Design of smart energy meter for the smart grid,} where a smart greed is a next generation power grid 
    having a two-way flow of electricity and information, see \cite{yan2013survey} for more details on smart grids.\\
    \cite{zheng2013smart} presented an overview of typical smart meter's aspects and functions  wrt. smart grid aspects.\\  \cite{kuzlu2014communication} analysed communication network requirements for smart grid applications.\\
    \cite{yaacoub2014automatic} proposed an approach on automatic meter reading in the smart grid using contention based random access over the free cellular spectrum.\\ 
    \cite{arif6529714} conducted a study on design and development of smart energy meter for the smart grid. 
    \item
    \emph{Privacy and security aspects  of smart meters} 
    are studied especially intensively over the last years, as the privacy and security concerns provide one of the biggest obstacles for the (potential) users of smart meters.\\
    \cite{yan2013efficient} proposed a security protocol for advanced metering infrastructure in smart grid. 
    \cite{sankar2013smart} proposed a theoretical framework to analyse privacy aspects of smart meters.\\ 
    \cite{albert2013smart} and \cite{beckel2014revealing} discussed what the consumption patterns derived using the smart meters might say about the  consumers.\\ 
\cite{chen2013non} presented an approach for 
non-intrusive occupancy monitoring using smart meters, having a goal to implement energy-efficiency optimizations based on the information of home's occupancy. Other approaches for occupancy detection from electricity consumption data were proposed by \cite{kleiminger2013occupancy,yang2014inferring,masoudifar2014monitoring,chen2018building} and \cite{tang2015meter}.
\\
\cite{tan2013increasing} proposed a solution to increase the
smart meter privacy through energy harvesting and storage devices.\\
\cite{eibl2015influence} analysed the influence of data granularity on smart meter privacy as well as what granularity should be used to prevent the interference of personal data  from load profiles by using non-intrusive appliance load monitoring methods. 
Another approach for preventing occupancy detection from smart meters was proposes by   \cite{chen2014combined,chen2015preventing}.\\
\cite{eibl2015privacy} elaborated a set of use cases for Smart
Metering,  formulated in a way that is
suitable for the development of privacy enhancing technologies. \\
\cite{eibl2018unsupervised} also presented a study on holiday detection from energy consumption data based on low-resolution smart meter data.\\
\cite{burkhart2018detecting} conducted a study where   swimming pools were detected through their filter pumps in load data with the 15-minute granularity prescribed by the European Union for smart meters, which demonstrates how vulnerable the private information might be through access to the meter readings data.
\end{itemize}
%..................

%===============================================
\section{\uppercase{AWS Rekognition vs. Google Cloud Vision}}
\label{sec:study}

To implement the proposed system, the were selecting between two computer vision technologies, AWS Rekognition vs. Google Cloud Vision. In the below sections we present the methodology of the comparison as well as the details of the conducted study.

\subsection{Methodology}

Reading utility meters involves several challenges for application of computer vision technologies:
reflection from the meters' glass, 
 clipped digits,  
 additional text on the meter that does not belong to the actual meter reading, 
   blur, 
   noise, as well as cases, where a meter has digital representation style for some readings but dial representation for other. %  

Images for the evaluation data set were selected based on their ``uniqueness'' -- images with unique meters or images with unique lighting. A total of 30 images were selected. This set of images were duplicated and modified with various effects in order to test the limitations of the different technologies. These effects are:

%===============================================

\begin{figure*}[ht!] 
\begin{center}
%\centering
\includegraphics[scale=0.8]{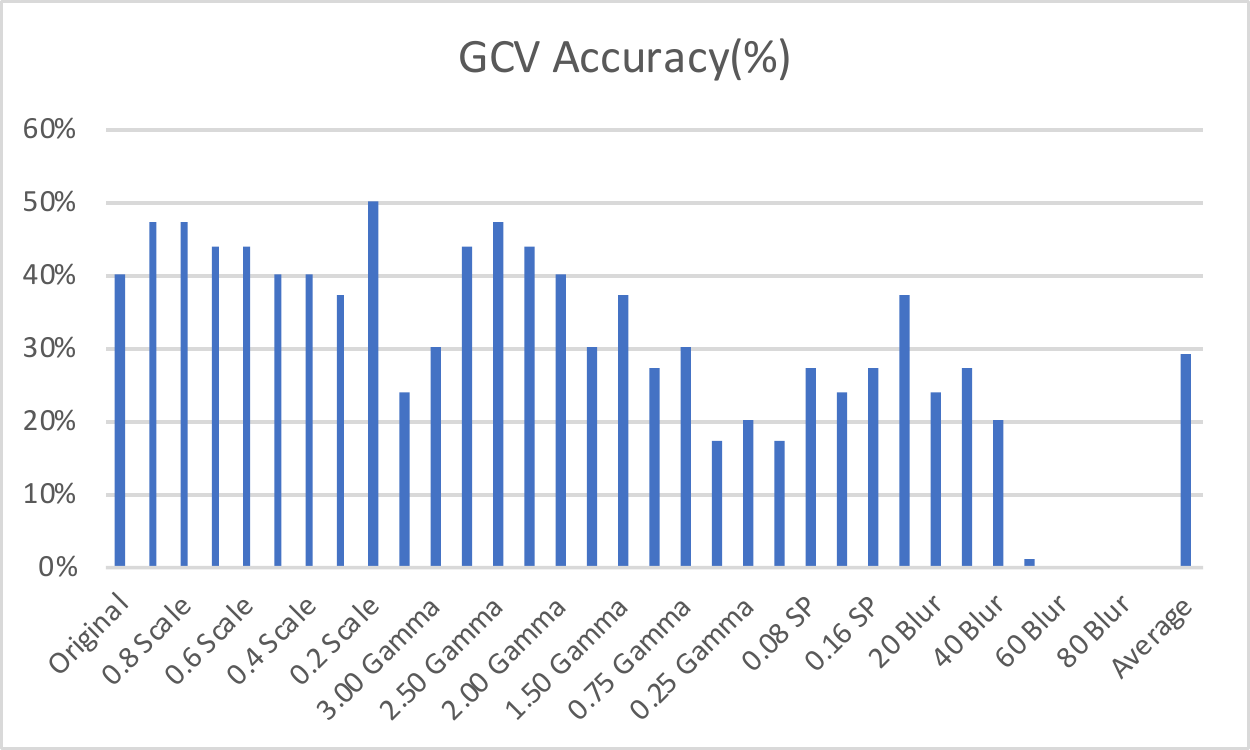}
\end{center}
\caption{Google Cloud Vision (GCV) Accuracy}
\label{fig:F1}
\end{figure*}

\begin{figure*}[ht!] 
\begin{center}
%\centering
\includegraphics[scale=0.8]{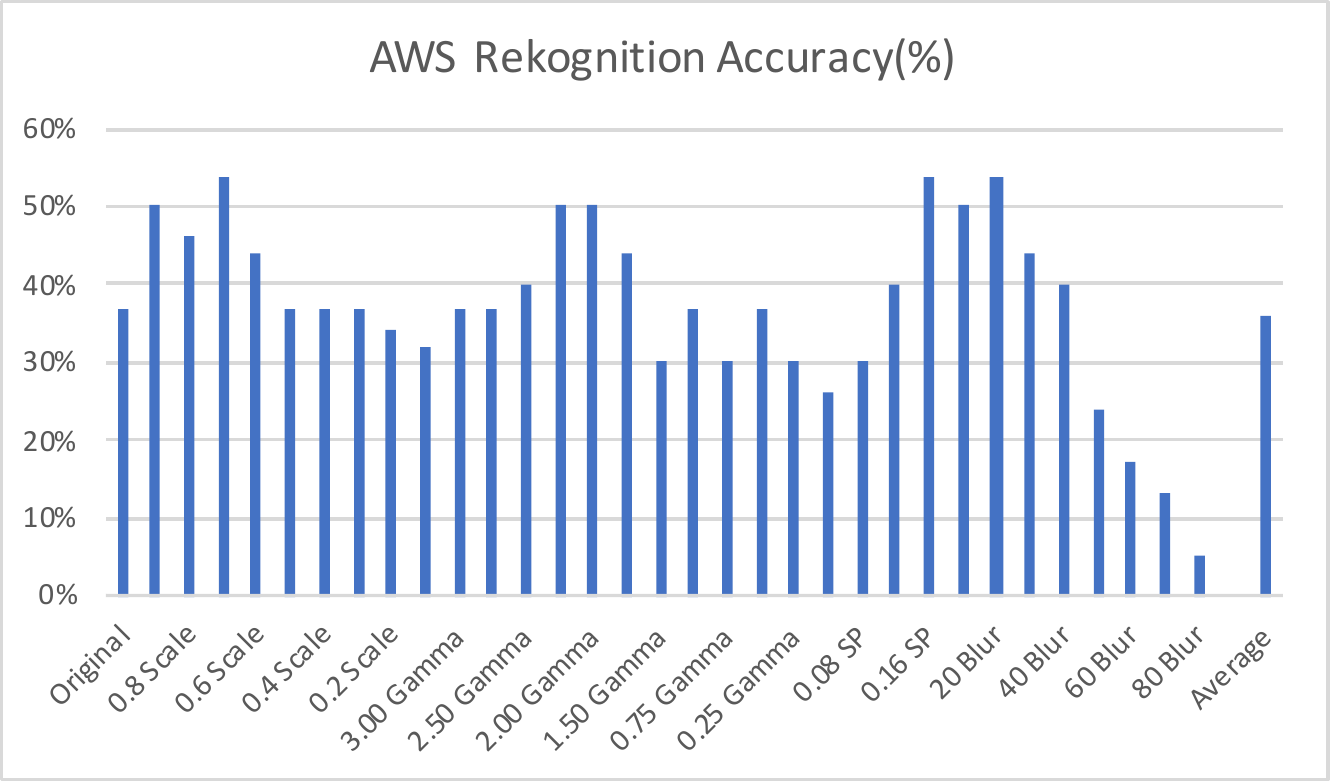}
\end{center}
\caption{AWS Recognition Accuracy}
\label{fig:F2}
\end{figure*}

\begin{itemize}
    \item \emph{Scaling:} The data set was scaled in steps of 0.1 ranging from a scale of 0.1 to 0.9 (10$\%$ to 90$\%$) of the original data set. 
    \item \emph{Blurring:} Blurring was done in steps of 10 from 10 to 90 with an open source blur algorithm that is based on the normalised box filter, see \cite{OpenCV}. The algorithm uses a normalised box filter, the numeral value adjusts the kernel size. Figures \ref{fig:blur}(a)--\ref{fig:blur}(c) present examples of blurring application with 30BLUR, 60BLUR, and 90BLUR, respectively.
    \item \emph{Gamma:} The gamma algorithm was used with an open source lookup table algorithm  \cite{OpenCV}.  The gamma correction to simulate different lightning conditions. Figures~\ref{fig:gamma}(a)--\ref{fig:gamma}(c) present examples of gamma algorithm application with 0.25GAMMA, 1.5GAMMA, and 3.0GAMMA, respectively.
    \item \emph{Noise:} The noise algorithm is based upon the salt and pepper noise algorithm that adds sharp and sudden disturbances in the image in the form of sparsely occurring white and black pixels, see \cite{Gonzalez:2001:DIP:559707}. This algorithm was included to further test the performance of the various technologies as noise arguably emulates ``dirt'' on meters. 
    %However, the noise applied is barely visible for a human eyes when images are down-scaled to fit into document. 
\end{itemize}

We calculated the accuracy of recognition 
calculated as the following simple formula (we measure the accuracy in percents, where $100\%$ means a totally accurate recognition):
\begin{equation}
 Accuracy   = \frac{CorrectResults}{Total} *100
 %(CorrectResults \div Total) *100
 \label{eq:e1}
\end{equation}
where\\ 
$CorrectResults$ is the number of results that  match with the original readings completely,\\
$Total$ presents the total number of images in data set. 
In our study, we had 30 images in each of the data sets. 

%===============================================
\subsection{{Results of the study}}

The results of the conducted study are summarised on Figures \ref{fig:F1} and \ref{fig:F2} for Google Cloud Vision and AWS Rekognition, respectively. 
The bar \emph{Original} presents the recognition results for the original data set. For this case,  Google Cloud Vision has performed slightly better than AWS Rekognition having a 3$\%$ higher accuracy.

\begin{figure*}[ht!] 
\begin{center}
%\centering
\includegraphics[scale=0.65]{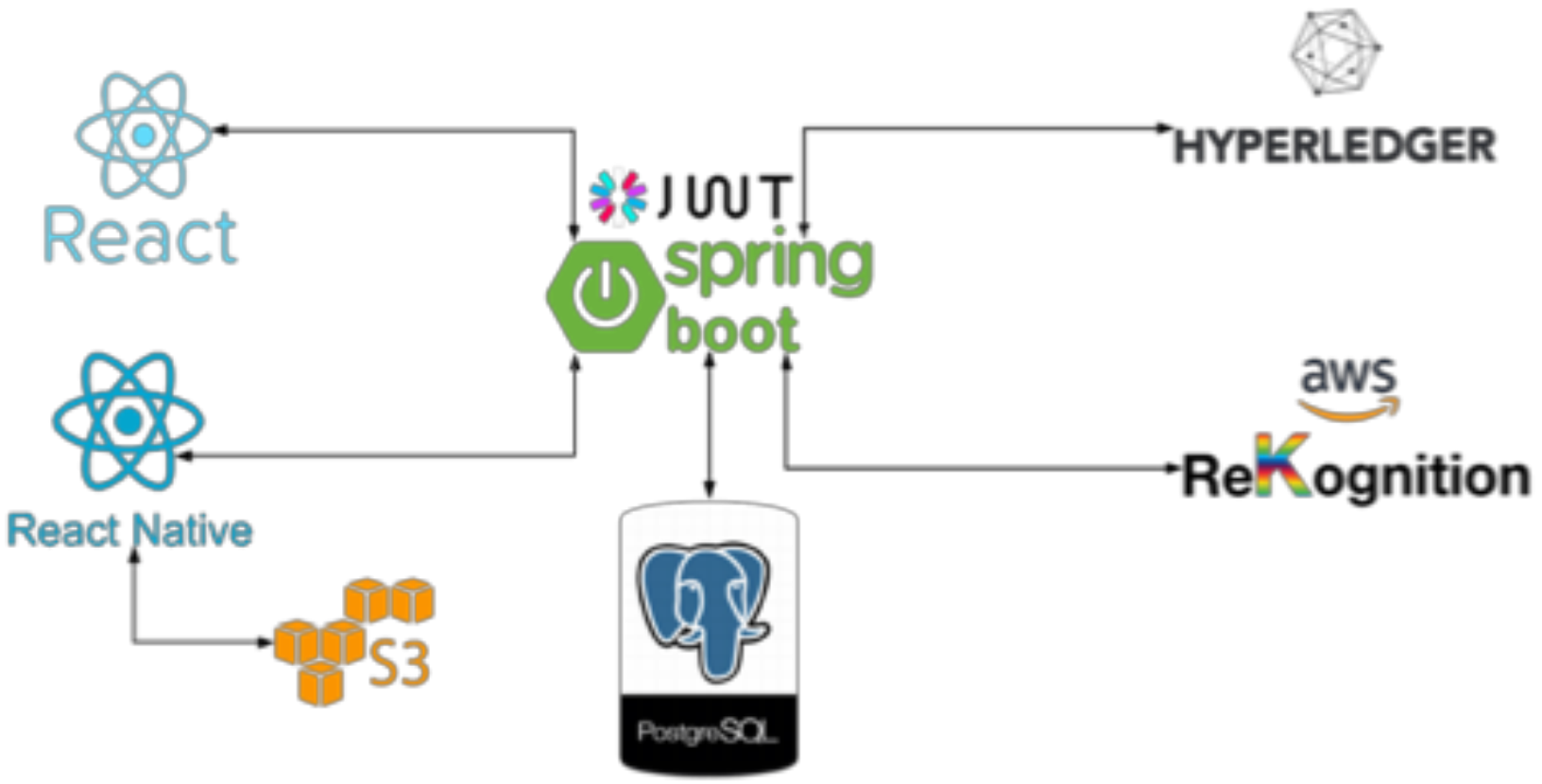}
\end{center}
\caption{Solution Architecture}
\label{fig:arch}
\end{figure*}

\emph{Scale Data set:}
There is a variation of 10$\%$ in the accuracy of the two models. AWS Rekognition has an overall higher efficiency than Google Cloud Vision with the former performing 10$\%$ better than the latter in every iteration. As the value of scaling is increased, accuracy is also increasing.

\emph{Gamma Dataset:} The variation between the two, in this case, is almost negligible, as both provide an accuracy of approx. 40$\%$.
SP Dataset: AWS Rekognition outperforms Google Cloud Vision with over 20$\%$ margin in accuracy. As the value of SP increases, so does the accuracy.

\emph{Blur Dataset:} This dataset proved to be a challenge for both the models, with AWS Rekognition reaching a top accuracy of 50$\%$ whereas the Google Cloud Vision only reached around 37$\%$ when blur level is 10. It dropped down to almost 0$\%$ when it reached around 40$\%$ blur in Google Cloud Vision and 90$\%$ blur in the case of AWS Rekognition. Even with higher blurred images, AWS Rekognition is able to detect some readings, unlike Google Cloud Vision where accuracy is 0$\%$.

Thus, on average, AWS Rekognition was able to perform approx. 7$\%$ better than Google Cloud Vision when same data set was provided.

%===============================================
\section{\uppercase{Proposed System}}
\label{sec:system}

Figure \ref{fig:arch} presents the solution architecture for the proposed system, where computer vision approaches are applied to capture meter readings using mobile phones. These readings should then be passed on to the core system to update consumer utility-charges accordingly. Consumers should then be able to view their renewed charges and usages in an internet browser.  
Thus, the mobile application is used to capture, upload and store an image of the meter to the system. 

The system will then analyse this image to identify meter readings and return the readings' values back to the user for confirmation. Once the user has confirmed the meter reading, it will be stored on a blockchain.

\begin{figure*}[ht!] 
\begin{center}
%\centering
\includegraphics[scale=0.7]{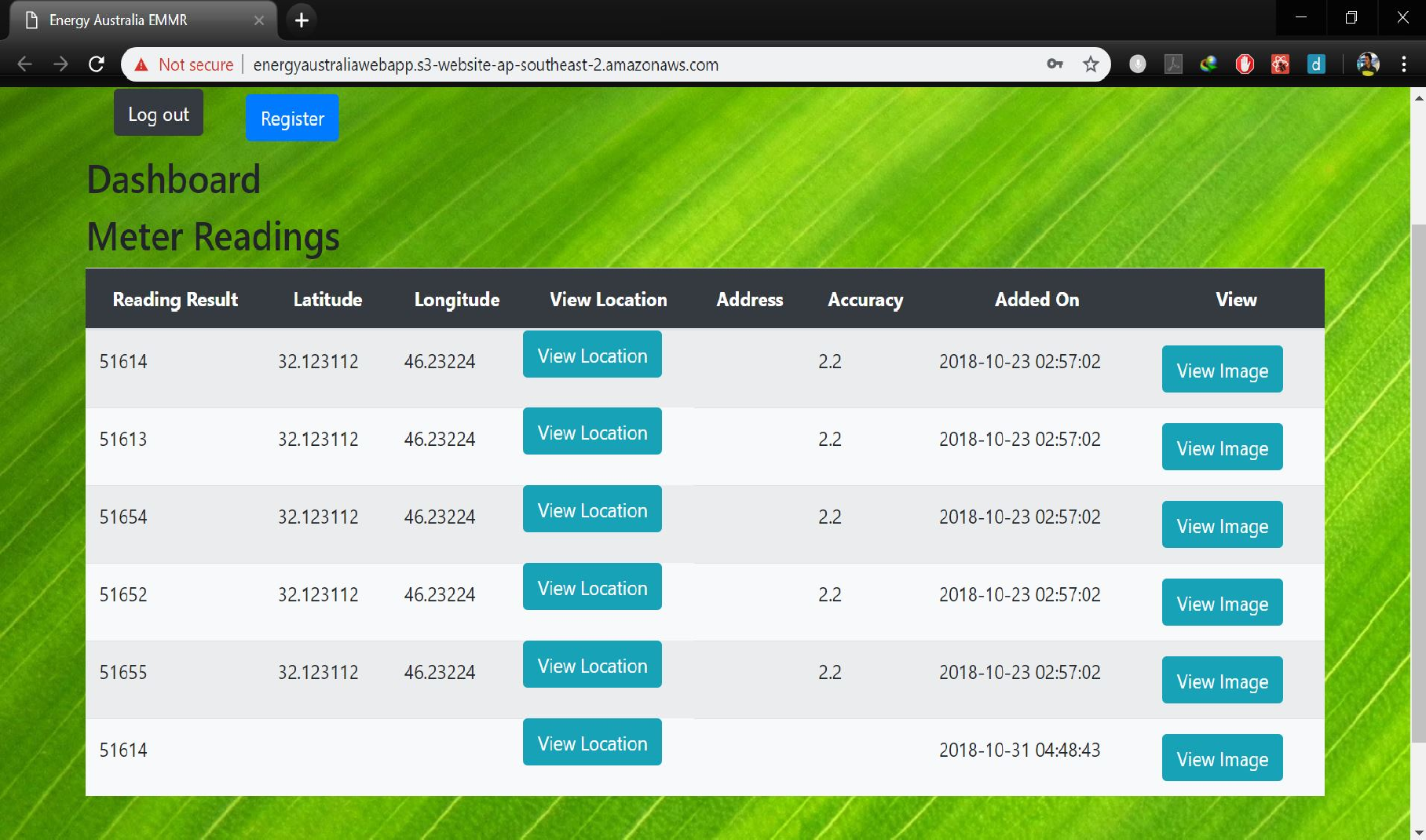}
\end{center}
\caption{Web Application (Admin): Meter reading results}
\label{fig:webapp2}
\end{figure*}

The proposed system has two core components providing  interfaces for two user types:  
\begin{itemize}
    \item an Android application developed for customers; the  application was built using React Native, which provides cross-platform compatibility between Android and iOS platforms (thus, development of an iOS version of the app will be less time-consuming);
    \item a Web application developed using ReactJS for admin users to audit the meter readings.
\end{itemize}
Mobile application and web application acts as a clients and call back-end APIs (application programming interfaces) running of Spring Boot. which is deployed on Amazon Web Services Elastic Beanstalk~\cite{awsEB}.  AWS Elastic Beanstalk reduces complexity without restricting choice or control, as it automatically handles the details of capacity provisioning, load balancing, scaling, and application health monitoring.

\begin{figure}[ht!] 
\begin{center}
%\centering
\includegraphics[scale=1]{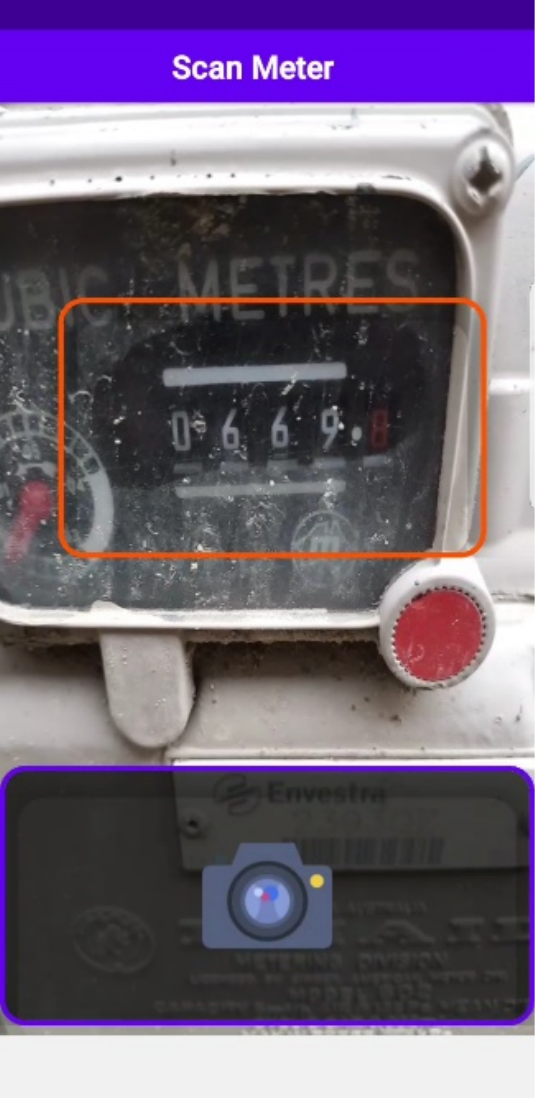}
\end{center}
\caption{Mobile Application (Customer View): Capturing an image of a meter}
\label{fig:app1}
\end{figure}

An example of a Web application page is presented in Figures~\ref{fig:webapp2}. Figure~\ref{fig:app1}  presents an examples of the mobile application pages. 

Spring Boot APIs are secured using JSON Web Token OAuth 2.0 security. The back-end uses PostgreSQL and Hyperledger Blockchain\footnote{%
\url{https://www.hyperledger.org}} to store data. Amazon Web Services (AWS) Rekognition is used to get the meter reading from the meter image. 
The choice of the computer vision technology is justified by the study presented in
Section~\ref{sec:study}.

\begin{figure*}[ht!] 
\begin{center}
%\centering
\includegraphics[scale=0.8]{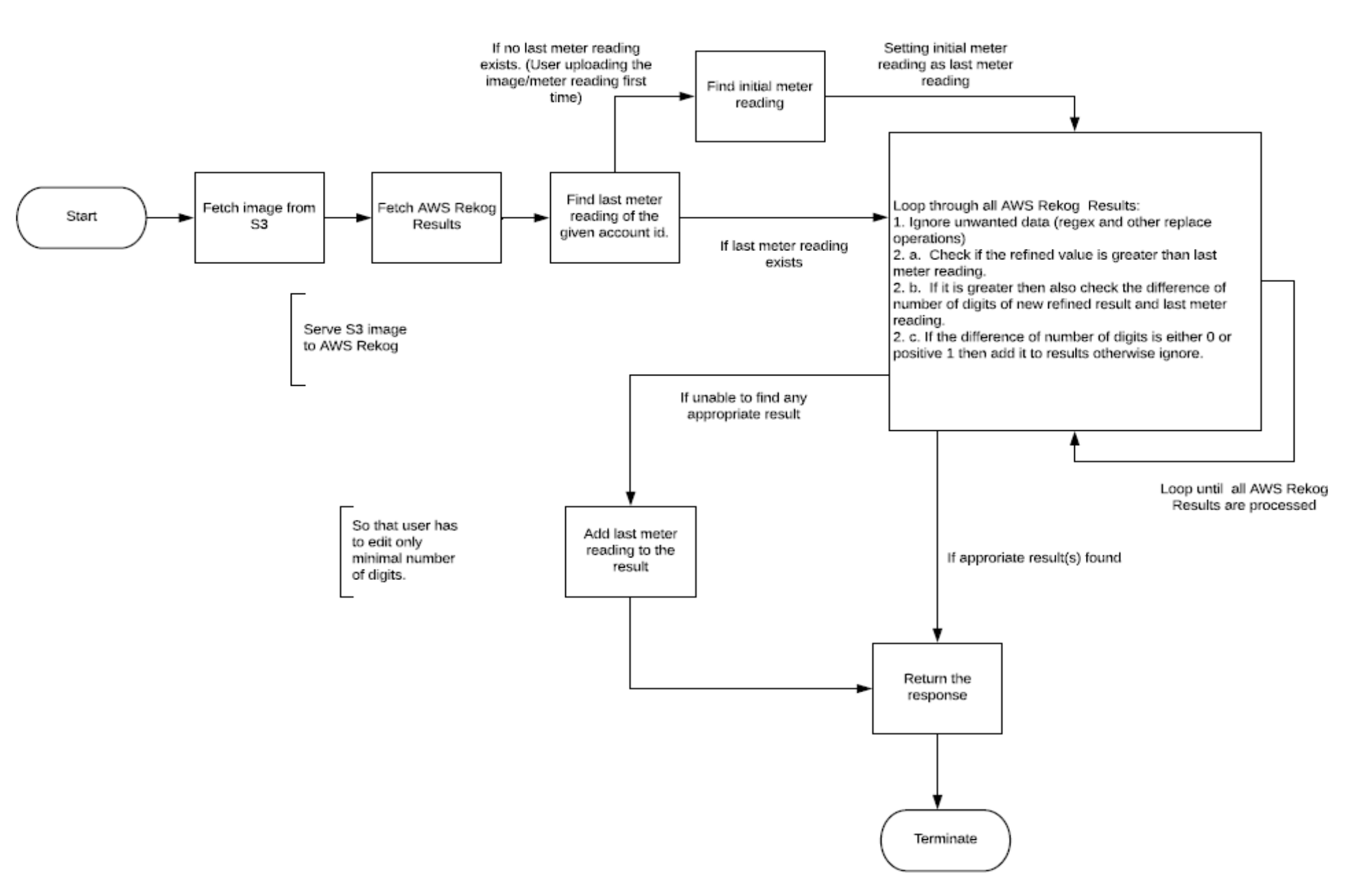}
\end{center}
\caption{Results Refinement Algorithm for Image Recognition}
\label{fig:alg1}
\end{figure*}

When a customer using the mobile application clicks an image of the meter (the application uses viewfinder technology as shown in Figure~\ref{fig:app1}), a Spring Boot API will be called to filter out the meter readings from the image and to forward the result to AWS Rekognition, which returns all the text at the Spring Boot level. Figure~\ref{fig:alg1} presents an algorithm we elaborated to filter out all irrelevant data and return only the relevant results back to the mobile application. The API takes the image URL and the storage bucket (S3) name from the client and returns the meter reading. Firstly, image is fetched from the URL and the bucket name, then the image is passed to the AWS Rekognition library, which is applied to identify all the text on the image. The algorithm further filters out all irrelevant text by considering the user's last meter reading or the initial meter reading, which was added to the system when the corresponding account was created. If the algorithm unable to return the scanned meter reading, it simply returns the last meter reading to the user, so that user has to change only the minimal number of digits. 

If the customer is satisfied with the image recognition results, the customer submits the meter reading, thus, another API will be called which stores the immutable data into Blockchain and mutable data into PostgreSQL database. The administrator can use the Web application to audit the meter readings at any time. Web application also calls Spring Boot APIs to get all customer details and their meter readings.

The blockchain also contains an interface from which the cloud-system can interact with.
The cloud-system provides a portal for administrators, where they can review customer meter readings through displaying previously uploaded images along with their respective geo-location coordinates. These features provide Energy Australia with a manual method of detecting falsified readings.

\begin{figure*}[ht!] 
\begin{center}
%\centering
\includegraphics[scale=0.9]{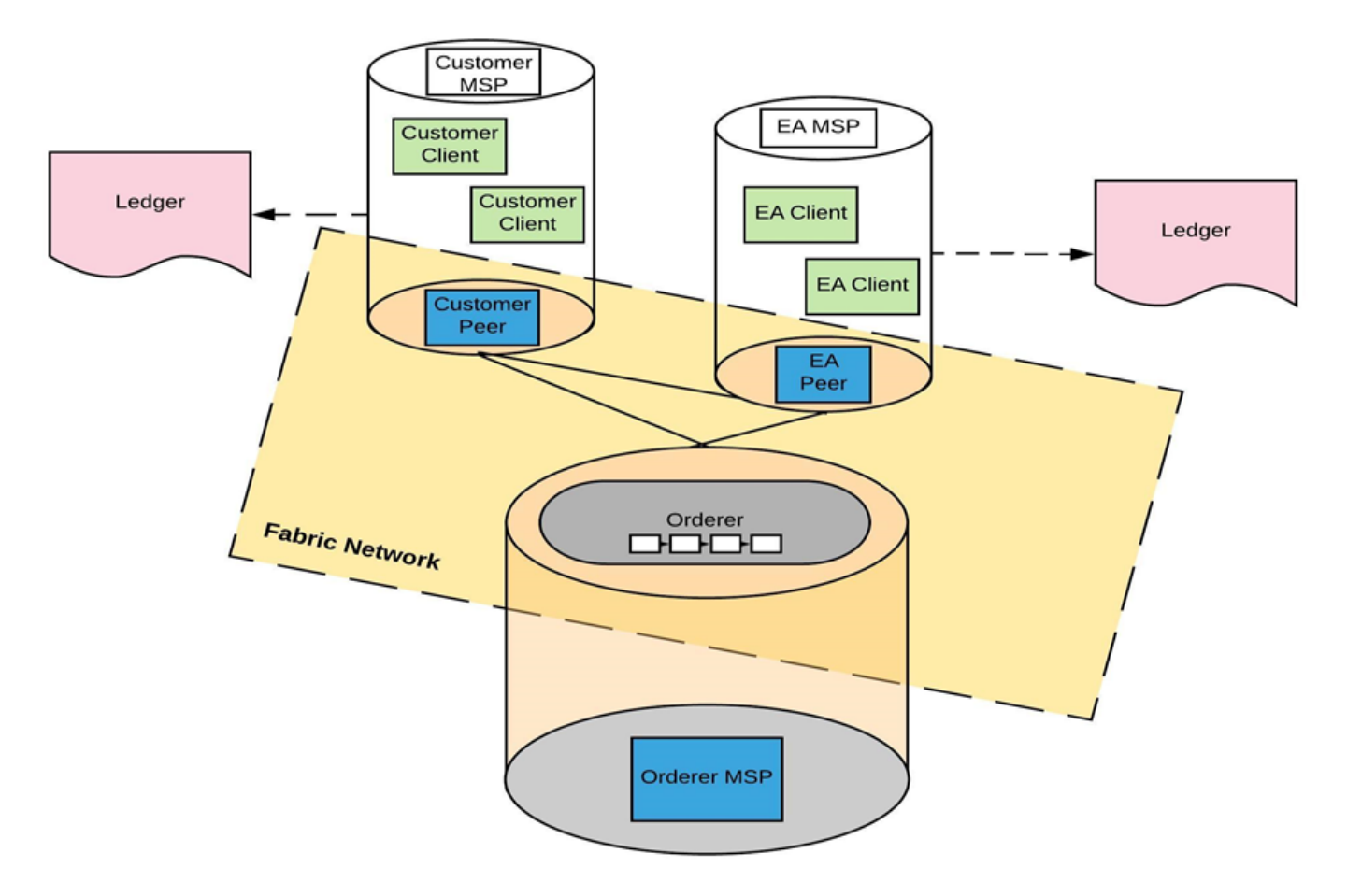}
\end{center}
\caption{Blockchain Architecture}
\label{fig:bc1}
\end{figure*}

The blockchain component consists of three nodes, see Figure~\ref{fig:bc1}: Customer Node, EA (Energy Australia) Node and Orderer Node; deployed using docker containers on three individual EC2 instances running on Ubuntu 16.04 Xenial Xerus. The peers are part of the Fabric and represent the node on the blockchain. Each Node has its own version of the Ledger using LevelDB. Each node also consists of MSP (Membership service provide) docker container used to provide signatures and certificates to new joining entities. Node.js is used on all the instances to expose the APIs for backend to interact with the Network.

When an update is made to the meter reading by a customer, it is sent by the customer node to the channel for verification. The EA node in this case acts as an endorser to verify the validity of the transaction. The requested transaction is executed on the endorsers’ version of the ledger. Once it is successful, the transaction for meter reading update is signed and sent back to the customer node. This signed transaction is then sent to Orderer. Orderer will verify the endorsed signature and wait for the next block to come up. Once a block is available it will update the meter reading and attach this block to the ledger. The block is then sent to all the nodes for inclusion in the Ledger.

Docker\footnote{\url{https://www.docker.com}} containers were used to launch the instances on to AWS EC2 instances. In this case, a docker container consist of six docker images: for Customer, for EA (Energy Australia), for Orderer, for Chaincode, for EAMSP (Energy Australia Membership Service Provider) and for Customer Membership Service Provider. The Chaincode docker consists of the channel on which the nodes are interacting and the latest version of Chaincode installed and instantiated. A simple web page is hosted to display the amount of transaction that have been committed to the ledger along with other network specifications. 
A shell bash script was written for each AWS EC2 instance to quickly generate all the artefacts required for Blockchain, to quickly setup and tear down the network for testing and development and finally for deployment. 
%

%===============================================
\section{\uppercase{Conclusions}} 
\label{sec:conclusions}

In this paper, we presented the core results of a research project conducted in collaboration with Energy Australia, an Australian electricity and gas retailing  company. 
The goal of the project was to provide a convenient alternative method for their current meter reading updating system focusing on non-smart meters. 
We implemented the proposed system as a cloud-based solution that applies
\begin{itemize}
    \item computer-vision technology to identify the meter readings automatically,
    \item blockchain technology to store the meter reading securely.
\end{itemize} 
We conducted a study to compare two computer vision technologies, 
Google Cloud Vision and AWS Rekognition,  applied for recognition in utility meter readings. The study demonstrated that AWS Rekognition provides better results for our application domain. Thus, AWS Rekognition was applied within the proposed system.

The developed system has two interfaces: 
\begin{itemize}
    \item the customer interface: a mobile application for automated capturing meter readings and managing the account details, such as customer's address, contact details, as well as the core details on the electricity and gas meters belonging to the customer;
    \item the administrator interface: a web application for management customers' accounts, details on the electricity and gas meters (including geo-location of the meters), as well as the stored images of the meter readings.
\end{itemize}

\emph{Future Work:}
As a possible future work direction we consider   to investigate further computer vision technologies, as the average accuracy values of Google Cloud Vision  and AWS  Rekognition applied for recognition in utility meter readings were not high.  
We consider to conduct a study to analyse the following technologies, also applied for recognition in utility meter readings:
    an open-source Tensorflow technique \cite{abadi2016tensorflow,abadi2017computational}
and 
    a commercial solution Anyline.  

%===============================================
\section*{{Acknowledgements}}
We would like to thank  Shine Solutions Group Pty Ltd for sponsoring this project under the research grant 
RE-03615. We also would like to thank Energy Australia for collaboration in this project. 
We also would like to thank the experts from the Shine Solutions Group, 
especially Aaron Brown and Alan Young for numerous discussions as well as their valuable advice and feedback.

\bibliographystyle{apalike}
{\small
%\bibliography{biblio}

}

\vfill
\end{document}